\documentclass[12pt]{article}

\usepackage{amssymb}
\usepackage{amsmath}
\usepackage{amscd}
\usepackage{mathrsfs}
\usepackage{amssymb}
\usepackage{feynmf}
\usepackage{graphicx}

\usepackage{tikz}

\def\dj{\hbox{d\kern-0.347em \vrule width 0.3em height 1.252ex depth
-1.21ex \kern 0.051em}}

\numberwithin{equation}{section}

\begin{document}

\setlength{\oddsidemargin}{0cm}
\setlength{\baselineskip}{7mm}


\thispagestyle{empty}
\setcounter{page}{0}

\begin{flushright}

\end{flushright}

\vspace*{1cm}

\begin{center}
{\bf \Large Color-Kinematics Duality and the Regge Limit}

\vspace*{0.4cm}

{\bf \Large of Inelastic Amplitudes}

%
%

\vspace*{0.3cm}

\vspace*{1cm}

Agust\'{\i}n Sabio Vera$^{a,}$\footnote{\tt 
a.sabio.vera@gmail.com}, 
Eduardo Serna Campillo$^{b,}$\footnote{\tt eduardo.serna@usal.es}, Miguel \' A. V\'azquez-Mozo$^{b,a,}$\footnote{\tt 
Miguel.Vazquez-Mozo@cern.ch}

\end{center}

\vspace*{0.0cm}

\begin{center}
$^{a}${\sl Instituto de F\'{\i}sica Te\'orica UAM/CSIC \&
Universidad Aut\'onoma de Madrid\\
C/ Nicol\'as Cabrera 15, E-28049 Madrid, Spain
}

$^{b}${\sl Departamento de F\'{\i}sica Fundamental \& IUFFyM,
 Universidad de Salamanca \\ 
 Plaza de la Merced s/n,
 E-37008 Salamanca, Spain
  }
\end{center}

\vspace*{2.5cm}

\centerline{\bf \large Abstract}

We investigate tree-level five-point amplitudes in scalar-QCD expressed in terms of 
Sudakov variables and find the equivalent 
``gravitational" counterparts using the color-kinematics duality proposed by Bern, Carrasco, and Johannson. Taking the multi-Regge limit in the gravitational amplitudes, we show that those pieces 
in the coupling of two reggeized gravitons to one on-shell graviton directly stemming from the double copy of the vertex for two reggeized gluons to one on-shell gluon are universal and 
properly reproduced by the duality. 

\vspace*{0.5cm}

\noindent

\newpage

\setcounter{footnote}{0}

\begin{fmffile}{effecvert}

\section{Introduction}

The relation between scattering amplitudes in gauge and gravity theories has been the 
subject of extensive research in recent years (for reviews on the subject we refer the reader 
to~\cite{review_gauge_gravity}). 
The fact that in a certain sense gravity can be regarded as the ``square'' of a gauge theory is suggested by string theory where tree-level
graviton (closed string) amplitudes can be written as linear combinations of the product 
of gluon (open string) amplitudes, as found by Kawai, Lewellen, and Tye (KLT)
\cite{KLT}. In the field theory limit they translate into similar relations for the 
corresponding graviton and gluon amplitudes in quantum field theory~\cite{Dunbar:1994bn,Bern:1993wt,Bern:1991an}. 
Unfortunately, no similar factorization is known for string loop amplitudes. 

More recently, a surprising relation between gluon and graviton amplitudes, dubbed color-kinematics duality, 
has been found by Bern, Carrasco, and Johansson (BCJ) \cite{BCJ} (see also~\cite{Bern:2010ue,BDHK,Bern:2011ia,Tye:2010dd,Mafra:2011kj}). 
The tree-level $n$-gluon amplitudes have the general structure
\begin{eqnarray}
\mathcal{A}(1,\ldots,n)_{\rm tree}=g^{n-2}\sum_{i\in \Gamma}{C_{i}N_{i}\over \prod_{\alpha}s_{\alpha}},
\label{eq:N-gluon_amplitude_general}
\end{eqnarray} 
where the sum runs over all Feynman diagram 
topologies $\Gamma$, $C_{i}$ are their color factors, $N_{i}$ the corresponding numerators, and 
$s_{\alpha}$ the kinematic invariants associated to the internal propagators of each diagram. 
Choosing the phases of the color factors properly, they satisfy Jacobi identities of the form
\begin{eqnarray}
C_{i}+C_{j}+C_{k}=0
\label{eq:jacobi_identity_general}
\end{eqnarray}
for certain triplets of indices $(i,j,k)$.

The amplitude \eqref{eq:N-gluon_amplitude_general}
is invariant under generalized gauge transformations \cite{BCJ,BDHK} shifting the numerators in the form 
$N_{i}\rightarrow N_{i}+\Delta_{i}$, where $\Delta_{i}$ are functions of the momenta satisfying 
\begin{eqnarray}
\sum_{i\in T}{C_{i}\Delta_{i}\over \prod_{\alpha}s_{\alpha}}=0.
\end{eqnarray}
These are generalizations of the standard gauge transformations shifting the gluon 
polarization vectors $\varepsilon_{\mu}(k)\rightarrow \varepsilon_{\mu}(k)+\lambda \,  k_{\mu}$, with $\lambda$ being 
a function of the momenta.
This new freedom can be used to choose the numerators $N_{i}$ such that they replicate the Jacobi identities of
the corresponding color factors shown in Eq. \eqref{eq:jacobi_identity_general}, namely
\begin{eqnarray}
N_{i}+N_{j}+N_{k}=0,
\label{eq:jacobi_identity_num_general}
\end{eqnarray}
for the same triplets $(i,j,k)$. The numerators obtained by the direct 
application of the Feynman rules in QCD  
do not fulfill these BCJ duality relations. In 
Ref.~\cite{BDHK} a nonlocal action for the gluon 
field was constructed whose Feynman rules give numerators satisfying \eqref{eq:jacobi_identity_num_general} 
for the five gluon tree-level amplitude.

Color-kinematics duality provides 
a prescription to construct the amplitude for the tree-level scattering 
of $n$ gravitons from two copies of the tree-level amplitude of $n$ gluons as
\cite{BCJ}
\begin{eqnarray}
-i
\mathcal{M}(1,\ldots,n)_{\rm tree}=\left({\kappa\over 2}\right)^{n-2}
\sum_{i\in \Gamma}{N_{i}\widetilde{N}_{i}\over \prod_{\alpha}s_{\alpha}}.
\end{eqnarray} 
Here $N_{i}$, $\widetilde{N}_{i}$ are two replicas of the numerators of the gauge theory amplitude 
\eqref{eq:N-gluon_amplitude_general} satisfying the BCJ duality 
relations~\eqref{eq:jacobi_identity_num_general} and $\kappa$ the gravitational coupling.
This hidden connection between gauge theories and gravity amplitudes is very remarkable since, unlike the KLT relations, 
it is expected to hold also for loop amplitudes before integration over the loop momenta.  

Whenever possible it is important to test the validity of this intriguing duality. Different kinematic limits of scattering amplitudes 
either in gauge or gravitational amplitudes are known which can serve as a test ground for the BCJ procedure. One recent example is the soft limit investigated in~\cite{Oxburgh:2012zr}. In the present work 
we focus on the Regge limit, which is well understood both in the gauge~\cite{BFKL1,BFKL2,BFKL3} and 
gravitational sides~\cite{Lipatov:2011ab,Lipatov:1982vv,Lipatov:1982it,Lipatov:1991nf,Bartels:2012ra,SVSCVM}.  At loop level, ${\cal N}$-supergravity amplitudes have been constructed from those of Yang-Mills theories even in the case of non-maximally supersymmetric theories~\cite{N=4CKD}. 
In~\cite{Bartels:2012ra} it has been shown that the four-graviton amplitudes at two loops calculated using the BCJ duality for  ${\cal N} \geq 4$ generate the correct Regge limit even at double logarithmic (in energy) accuracy, which goes beyond the well-known  
exponentiation of infrared divergencies. In the same work predictions have been made for graviton scattering in all supergravities and 
Einstein-Hilbert gravity to arbitrary order in the gravitational coupling  which should serve as a good test of the color-kinematics 
duality at higher orders. 

In Ref.~\cite{SVSCVM} we carried out a study of exact inelastic amplitudes both in QCD and Einstein-Hilbert 
gravity. Using a representation in Sudakov variables, we were able to reproduce the results of Lipatov for the emission of a 
graviton in multi-Regge kinematics (MRK). It turns out that this limit can be related to that of gluon production in QCD and 
an interesting double-copy structure emerges, which we investigate in the present work at the light of the color-kinematics duality. Here we find it useful to investigate the color-kinematics duality in scalar QCD (sQCD) by studying the scattering of two distinct scalars with the associated emission of a gluon in MRK, in this way it is possible to understand which pieces are somehow universal in the Regge limit 
and which ones are dependent on the matter content of the gauge theory when applying the BCJ procedure. 

\section{sQCD and color-kinematics duality}

Our starting point is the tree-level contribution to the scattering of two distinguishable massless scalar particles with the 
emission of one gluon in sQCD
\begin{eqnarray}
\Phi(p,j)+\Phi'(q,m)\longrightarrow \Phi(p',i)+\Phi(q',i)+g(k,a,\varepsilon),
\end{eqnarray}
where in brackets we indicated the momenta and color quantum numbers of the involved particles, as well as 
the polarization vector in the case of the gluon. We choose this amplitude in order 
to reduce the number of diagrams to be calculated and to understand how 
the BCJ procedure fails when in the external states we not only have gluons but 
also scalar fields.  The amplitude 
receives contributions from the following seven diagrams 
\begin{eqnarray}
\nonumber \\[0.2cm]
\mathcal{A}&\equiv&\hspace*{1.5cm}
\parbox{42mm}{
\begin{fmfgraph*}(90,60)
\fmfbottom{i1,d1,o1}
\fmftop{i2,d2,o2}
\fmfright{r1}
\fmflabel{$(p,j)$}{i2}
\fmflabel{$(p',i)$}{o2}
\fmflabel{$(k,a)$}{r1}
\fmflabel{$(q,n)$}{i1}
\fmflabel{$(q',m)$}{o1}
\fmf{plain}{i1,v1,v2,o1}
\fmf{plain}{i2,v3,v4,o2}
\fmf{phantom}{i1,o1}
\fmf{gluon,tension=0}{v1,v3}
\fmf{gluon,tension=0}{v4,r1}
\end{fmfgraph*} 
}\hspace*{0.5cm}+\hspace*{1.4cm}
\parbox{42mm}{
\begin{fmfgraph*}(90,60)
\fmfbottom{i1,d1,o1}
\fmftop{i2,d2,o2}
\fmfright{r1}
\fmflabel{$(p,j)$}{i2}
\fmflabel{$(p',i)$}{o2}
\fmflabel{$(k,a)$}{r1}
\fmflabel{$(q,n)$}{i1}
\fmflabel{$(q',m)$}{o1}
\fmf{plain}{i1,v1,v2,o1}
\fmf{phantom}{i1,o1}
\fmf{plain}{i2,v3,v4,o2}
\fmf{gluon,tension=0}{v2,v4}
\fmf{gluon,tension=0,rubout}{r1,v3}
\end{fmfgraph*} 
} \nonumber \\[0.9cm]
& & \hspace*{-0.4cm} +\hspace*{1.5cm}
\parbox{42mm}{
\begin{fmfgraph*}(90,60)
\fmfbottom{i1,d1,o1}
\fmftop{i2,d2,o2}
\fmfright{r1}
\fmflabel{$(p,j)$}{i2}
\fmflabel{$(p',i)$}{o2}
\fmflabel{$(k,a)$}{r1}
\fmflabel{$(q,n)$}{i1}
\fmflabel{$(q',m)$}{o1}
\fmf{plain}{i1,v2,o1}
\fmf{phantom}{i1,o1}
\fmf{plain}{i2,v4,o2}
\fmf{gluon,tension=0}{v2,v4}
\fmf{gluon,tension=0,rubout}{v4,r1}
\end{fmfgraph*} 
}\hspace*{0.5cm}+\hspace*{1.4cm}
\parbox{42mm}{
\begin{fmfgraph*}(90,60)
\fmfbottom{i1,d1,o1}
\fmftop{i2,d2,o2}
\fmfright{r1}
\fmflabel{$(p,j)$}{i2}
\fmflabel{$(p',i)$}{o2}
\fmflabel{$(k,a)$}{r1}
\fmflabel{$(q,n)$}{i1}
\fmflabel{$(q',m)$}{o1}
\fmf{plain}{i1,v1,v2,o1}
\fmf{phantom}{i1,o1}
\fmf{plain}{i2,v3,v4,o2}
\fmf{gluon,tension=0}{v1,v3}
\fmf{gluon,tension=0,rubout}{v2,r1}
\end{fmfgraph*} 
}
\label{eq:feyn_diag_scalars_gauge}
\end{eqnarray}
\begin{eqnarray}
\nonumber \\[0.9cm]
& & \hspace*{-0.4cm} +\hspace*{1.5cm}
\parbox{42mm}{
\begin{fmfgraph*}(90,60)
\fmfbottom{i1,d1,o1}
\fmftop{i2,d2,o2}
\fmfright{r1}
\fmflabel{$(p,j)$}{i2}
\fmflabel{$(p',i)$}{o2}
\fmflabel{$(k,\epsilon)$}{r1}
\fmflabel{$(q,n)$}{i1}
\fmflabel{$(q',m)$}{o1}
\fmf{plain}{i1,v1,v2,o1}
\fmf{phantom}{i1,o1}
\fmf{plain}{i2,v3,v4,o2}
\fmf{gluon,tension=0}{v2,v4}
\fmf{gluon,tension=0,rubout}{v1,r1}
\end{fmfgraph*} 
}\hspace*{0.5cm}+\hspace*{1.4cm}
\parbox{42mm}{
\begin{fmfgraph*}(90,60)
\fmfbottom{i1,d1,o1}
\fmftop{i2,d2,o2}
\fmfright{r1}
\fmflabel{$(p,j)$}{i2}
\fmflabel{$(p',i)$}{o2}
\fmflabel{$(k,a)$}{r1}
\fmflabel{$(q,n)$}{i1}
\fmflabel{$(q',m)$}{o1}
\fmf{plain}{i1,v2,o1}
\fmf{phantom}{i1,o1}
\fmf{plain}{i2,v4,o2}
\fmf{gluon,tension=0}{v2,v4}
\fmf{gluon,tension=0,rubout}{v2,r1}
\end{fmfgraph*} 
}
\nonumber \\[0.9cm]
& & 
\hspace*{2.5cm}
+\hspace*{1cm}
\parbox{42mm}{
\begin{fmfgraph*}(90,60)
\fmfbottom{i1,d1,o1}
\fmftop{i2,d2,o2}
\fmfright{r1}
\fmflabel{$(p,j)$}{i2}
\fmflabel{$(p',i)$}{o2}
\fmflabel{$(q,n)$}{i1}
\fmflabel{$(q',m)$}{o1}
\fmf{plain}{i1,v1,o1}
\fmf{plain}{i2,v3,o2}
\fmf{gluon,tension=0}{v1,v3}
\fmf{phantom}{i1,o1}
\end{fmfgraph*} 
}\hspace*{-2.4cm}
\parbox{42mm}{
\begin{fmfgraph*}(40,20)
\fmfleft{i1}
\fmfright{o1}
\fmflabel{$(k,a)$}{o1}
\fmf{gluon,tension=1}{i1,o1}
\end{fmfgraph*} 
}
\hspace*{-2cm} \nonumber
\\ \nonumber
\end{eqnarray}
and has the general structure
\begin{eqnarray}
\mathcal{A}=g^{3}\left({C_{1}N_{1}\over t's_{p'k}}+{C_{2}N_{2}\over t's_{pk}}+{C_{3}N_{3}\over t'}
+{C_{4}N_{4}\over t s_{q'k}}+{C_{5}N_{5}\over t s_{qk}}+{C_{6}N_{6}\over t}+{C_{7}N_{7}\over tt'}\right),
\end{eqnarray}
where the numbering corresponds to the order in which the diagrams appear in Eq. \eqref{eq:feyn_diag_scalars_gauge}.
In writing the amplitude we have introduced the following kinematic invariants
\begin{eqnarray}
t&=& (p-p')^{2}, \nonumber \\
t'&=& (q-q')^{2}, \nonumber \\
s_{pk}&=& (p+k)^{2}, \nonumber \\
s_{p'k}&=& (p'+k)^{2},  \\
s_{qk} &=& (q+k)^{2}, \nonumber \\
s_{q'k}&=& (q'+k)^{2}, \nonumber
\end{eqnarray}
and color factors
\begin{eqnarray}
C_{1}&=& T^{a}_{ik}T^{b}_{kj}T^{b}_{mn},\nonumber \\
C_{2}&=& T^{b}_{ik}T^{a}_{kj}T^{b}_{mn},\nonumber \\
C_{3}&=& T^{a}_{ik}T^{b}_{kj}T^{b}_{mn}+T^{b}_{ik}T^{a}_{kj}T^{b}_{mn} ,\nonumber \\
C_{4}&=& T^{b}_{ij}T^{a}_{mk}T^{b}_{kn}, \label{eq:color_factors_scalar}\\
C_{5}&=& T^{b}_{ij}T^{b}_{mk}T^{a}_{kn},   \nonumber \\
C_{6}&=&  T^{b}_{ij}T^{a}_{mk}T^{b}_{kn}+T^{b}_{ij}T^{b}_{mk}T^{a}_{kn},         \nonumber \\
C_{7}&=& if^{abc}T^{b}_{ij}T^{c}_{mn}  ,       \nonumber
\end{eqnarray}
with $T^{a}_{ij}$ being the generators of the representation in which the scalar fields transform. They
satisfy the Jacobi-like identities
\begin{eqnarray}
C_{1}-C_{2}+C_{7}&=&0,\nonumber \\
C_{1}+C_{2}-C_{3}&=&0,\nonumber \\
C_{4}-C_{5}-C_{7}&=&0, \label{eq:color_jacobi_id}\\
C_{4}+C_{5}-C_{6}&=&0.
\nonumber
\end{eqnarray}
Applying the Feynman rules\footnote{See, {\it e.g.}, \cite{itzykson_zuber}. Notice, however, that in the conventions of
this reference the gauge group generators are anti-Hermitian, whereas we take them Hermitian.} 
of sQCD, we find the following form for the numerators $N_{i}$
\begin{eqnarray}
\label{coeficientes}
N_{1} &=& 2i[(p+p'+k)\cdot (q+q')][p'\cdot\varepsilon(k)], \nonumber \\
N_{2} &=& -2i[(p+p'-k)\cdot (q+q')][p\cdot \varepsilon(k)], \nonumber \\
N_{3} &=& -i(q+q')\cdot\varepsilon(k),\nonumber \\
N_{4} &=& 2i[(q+q'+k)\cdot (p+p')][q'\cdot \varepsilon(k)], \\
N_{5} &=& -2i[(q+q'-k)\cdot (p+p')][q\cdot \varepsilon(k)], \nonumber \\
N_{6} &=& -i(p+p')\cdot\varepsilon(k),   \nonumber \\
N_{7} &=&  -i\Big\{[(q+q')\cdot (p-p'+k)][(p+p')\cdot\varepsilon(k)] \nonumber \\
&+&[(p+p')\cdot (q+q')][(p'-p-q'+q)\cdot\varepsilon(k)]
\nonumber \\
&+&[(p+p')\cdot(q'-q-k)][(q+q')\cdot\varepsilon(k)]\Big\}.           
\nonumber
\end{eqnarray}
These numerators do not satisfy the BCJ duality relations derived from 
Eq.~\eqref{eq:color_jacobi_id} and are therefore not ready to apply the color-kinematic 
duality prescription. The generalized nonlocal gauge transformation
\begin{eqnarray}
N_{1}'&=& N_{1}+s_{p'k}\left(N_{3}-{N_{7}\over 2t}\right), \nonumber \\
N_{2}'&=& N_{2}+s_{pk}\left(N_{3}+{N_{7}\over 2t}\right), \nonumber \\
N_{3}'&=& N_{3},\nonumber \\
N_{4}'&=& N_{4}+s_{q'k}\left(N_{6}+{N_{7}\over 2t'}\right), \\
N_{5}'&=& N_{5}+s_{qk}\left(N_{6}-{N_{7}\over 2t'}\right), \nonumber \\
N_{6}'&=& N_{6}, \nonumber \\
N_{7}'&=& N_{7}, \nonumber
\end{eqnarray}
recasts the amplitude in terms of only four color factors
\begin{eqnarray}
\mathcal{A}=g^{3}\left({C_{1}N_{1}'\over t's_{p'k}}+{C_{2}N_{2}'\over t's_{pk}}
+{C_{4}N_{4}'\over t s_{q'k}}+{C_{5}N_{5}'\over t s_{qk}}\right),
\end{eqnarray}
satisfying the single Jacobi identity [derived from Eq.~\eqref{eq:color_jacobi_id}]
\begin{eqnarray}
C_{1}-C_{2}+C_{4}-C_{5}=0.
\end{eqnarray}
These new numerators $N_{i}'$ do not comply with BCJ duality. 
To fix this we perform a further generalized gauge transformation of the form
\begin{eqnarray}
N_{1}''&=& N_{1}'+\alpha t's_{p'k}, \nonumber \\
N_{2}''&=& N_{2}'-\alpha t's_{pk},\nonumber \\
N_{4}''&=& N_{4}+\alpha t s_{q'k}, \\
N_{5}''&=& N_{5}-\alpha t s_{qk},\nonumber
\end{eqnarray}
where the function $\alpha$ is determined by requiring 
\begin{eqnarray}
N_{1}''-N_{2}''+N_{3}''-N_{4}''=0.
\label{eq:BCJ_dualityN1-N5}
\end{eqnarray}
This gives
\begin{eqnarray}
\alpha={-N_{1}'+N_{2}'-N_{4}'+N_{5}'\over t'(s_{pk}+s_{p'k})+t(s_{qk}+s_{q'k})}.
\end{eqnarray}

After all these algebraic 
manipulations we have managed to write our amplitude in the form
\begin{eqnarray}
\mathcal{A}=g^{3}\left({C_{1}N_{1}''\over t's_{p'k}}+{C_{2}N_{2}''\over t's_{pk}}
+{C_{4}N_{4}''\over t s_{q'k}}+{C_{5}N_{5}''\over t s_{qk}}\right),
\end{eqnarray}
where the numerators satisfy~\eqref{eq:BCJ_dualityN1-N5}. 
The next step is to apply the color-kinematics duality
prescription to construct the gravitational amplitude, {\it i.e.}
\begin{eqnarray}
-i\mathcal{M}=\left({\kappa\over 2}\right)^{3}\left({N_{1}''\widetilde{N}_{1}''\over t's_{p'k}}
+{N_{2}''\widetilde{N}_{2}''\over t's_{pk}}+{N_{4}''\widetilde{N}_{4}''\over t s_{q'k}}
+{N_{5}''\widetilde{N}_{5}''\over t s_{qk}}\right).
\label{eq:candidate_grav_amplitude}
\end{eqnarray}
The graviton polarization tensor $\epsilon_{\mu\nu}(k)$ is reconstructed as
\begin{eqnarray}
\varepsilon_{\mu}(k)\widetilde{\varepsilon}_{\nu}(k)\longrightarrow \epsilon_{\mu\nu}(k),
\end{eqnarray}
with $\varepsilon_{\mu}(k)$, $\widetilde{\varepsilon}_{\mu}(k)$ being the gluon polarization vectors contained in the
numerators $N''_{i}$ and $\widetilde{N}''_{i}$. 

At this point it is convenient to redefine the momenta according to
\begin{eqnarray}
p'=p-k_{1},\hspace*{1cm} q'=q+k_{2}, \hspace*{1cm} k=k_{1}-k_{2}
\end{eqnarray}
and write $k_{1}$ and $k_{2}$ using Sudakov parameters
\begin{eqnarray}
k_{1}^{\mu}=\alpha_{1}p^{\mu}+\beta_{1}q^{\mu}+k_{1,\perp}^{\mu},\hspace*{1cm}
k_{2}^{\mu}=\alpha_{2}p^{\mu}+\beta_{2}q^{\mu}+k_{2,\perp}^{\mu}.
\end{eqnarray} 
In this representation for the momenta, the amplitude 
\begin{eqnarray}
-i\mathcal{M}\equiv -iA_{kk}\mathcal{M}^{\mu\nu}\epsilon_{\mu\nu}(k)
\label{eq:pre-tensor_structure}
\end{eqnarray} 
can be shown to have the following tensor structure
\begin{eqnarray}
\mathcal{M}^{\mu\nu}&=&(k_{1}+k_{2})_{\perp}^{\mu}(k_{1}+k_{2})_{\perp}^{\nu}+\mathcal{A}_{kp}\Big[
(k_{1}+k_{2})_{\perp}^{\mu}p^{\nu}+p^{\mu}(k_{1}+k_{2})_{\perp}^{\nu}\Big] \nonumber \\
&+&\mathcal{A}_{kq}\Big[
(k_{1}+k_{2})_{\perp}^{\mu}q^{\nu}+q^{\mu}(k_{1}+k_{2})_{\perp}^{\nu}\Big]
+\mathcal{A}_{pq}\Big(p^{\mu}q^{\nu}+q^{\mu}p^{\nu}\Big)
\label{eq:tensor_structure}
 \\
&+& \mathcal{A}_{qq}q^{\mu}q^{\nu}+\mathcal{A}_{pp}p^{\mu}p^{\nu}. \nonumber
\end{eqnarray}
Let us point out that in Eq.~\eqref{eq:pre-tensor_structure}
we have factored out the coefficient $A_{kk}$ of the term proportional to
$(k_{1}+k_{2})_{\perp}^{\mu}(k_{1}+k_{2})_{\perp}^{\nu}$.
In the MRK limit
\begin{eqnarray}
1 \gg \alpha_{1} \gg \alpha_{2}, \hspace*{1cm} 1\gg |\beta_{2}|\gg |\beta_{1}|,
\end{eqnarray}
the prefactors $\mathcal{A}_{i}$ have the following form
\begin{eqnarray}
\mathcal{A}_{pp}&=& \left(\alpha_{1}-2{\beta_{1}\over \beta_{2}}\right)^{2}+2\alpha_{2}\beta_{1}\left(
{\alpha_{1}+\beta_{2}\over \beta_{2}^{2}}\right)+\ldots, \nonumber \\
\mathcal{A}_{qq}&=&\left(\beta_{2}+2{\alpha_{2}\over \alpha_{1}}\right)^{2}-2\alpha_{2}\beta_{1}\left(
{\alpha_{1}+\beta_{2}\over \alpha_{1}^{2}}\right)+\ldots, \nonumber \\
\mathcal{A}_{pq} &=& \left(\alpha_{1}-2{\beta_{1}\over \beta_{2}}\right)\left(\beta_{2}+2{\alpha_{2}\over \alpha_{1}}\right)
+\ldots,  
\label{eq:As_ckdual}\\
\mathcal{A}_{kp} &=& -\left(\alpha_{1}-2{\beta_{1}\over \beta_{2}}\right)+\ldots, \nonumber \\
\mathcal{A}_{kq} &=& -\left(\beta_{2}+2{\alpha_{2}\over \alpha_{1}}\right)+\ldots\nonumber
\end{eqnarray} 

We want to compare these results, obtained after applying the BCJ color-kinematics 
duality, with the ones found in Ref.~\cite{SVSCVM} using traditional Feynman rules for the gravitational scattering of two different scalars with emission of a graviton:
\begin{eqnarray}
\Phi(p)+\Phi'(q)\longrightarrow \Phi(p')+\Phi'(q')+G(k,\epsilon).
\end{eqnarray}
There we obtained that the tree-level amplitude
in the MRK regime can be written as [in the notation  of Eq.~(\ref{eq:pre-tensor_structure})]
\begin{eqnarray}
\mathcal{M}^{\mu\nu}=\Omega^{\mu}\Omega^{\nu}-\mathcal{N}^{\mu}\mathcal{N}^{\nu},
\label{eq:omega2-n2}
\end{eqnarray}
where
\begin{eqnarray}
\Omega^{\mu}&=&\left(\alpha_{1}-{2\beta_{1}\over \beta_{2}}\right)p^{\mu}+\left(\beta_{2}+{2\alpha_{2}\over \alpha_{1}}
\right)q^{\mu}-(k_{1}+k_{2})_{\perp}^{\mu}
\end{eqnarray}
is Lipatov's MRK effective vertex for the coupling of two reggeized gluons to an on-shell gluon in QCD, and
\begin{eqnarray}
\mathcal{N}^{\mu}&=&-2i\sqrt{\beta_{1}\alpha_{2}}\left({p^{\mu}\over \beta_{2}}+{q^{\mu}\over \alpha_{1}}\right).
\end{eqnarray}
The term $\mathcal{N}^{\mu}\mathcal{N}^{\nu}$ in Eq. \eqref{eq:omega2-n2} 
is responsible for the cancellation of overlapping singularities in simultaneous channels required by the Steinmann 
relations~\cite{Steinmann} (see also~\cite{Bartels:2008ce,Bartels:2008sc} for a recent discussion). 
It is important to realize that Eq.~\eqref{eq:As_ckdual} does not reproduce the full structure indicated in Eq.~\eqref{eq:omega2-n2}. 
This is an effect of applying the color-kinematics duality to an amplitude where some of the external states are not gluons in the 
gauge theory side. However, we can see that the term corresponding purely to the ``square" of the gauge vertex, $\Omega^{\mu}\Omega^{\nu}$, is correctly reproduced by the BCJ prescription in the Regge limit. 

In order to show that, although the $\mathcal{N}^{\mu}\mathcal{N}^{\nu}$ terms are not given by BCJ color-kinematic duality, 
the $\Omega^{\mu}\Omega^{\nu}$ contributions are indeed retrieved, 
we apply the prescription to a set of numerators satisfying different 
BCJ duality identities. Using the local generalized 
gauge transformation
\begin{eqnarray}
N_{1}'&=& N_{1}+s_{p'k}N_{3}, \nonumber \\
N_{2}'&=& N_{2}+s_{pk}N_{3}, \nonumber \\
N_{3}'&=& N_{3},  \\
N_{4}'&=& N_{4}+s_{q'k}N_{6}, \nonumber \\
N_{5}'&=& N_{5}+s_{qk}N_{6}, \nonumber \\
N_{7}'&=& N_{7},
\end{eqnarray}
the sQCD amplitude can be written in the form
\begin{eqnarray}
\mathcal{A}=g^{3}\left({C_{1}N_{1}'\over t's_{p'k}}+{C_{2}N_{2}'\over t's_{pk}}
+{C_{4}N_{4}'\over t s_{q'k}}+{C_{5}N_{5}'\over t s_{qk}}+{C_{7}N'_{7}\over tt'}\right).
\end{eqnarray}
A further transformation 
\begin{eqnarray}
N_{1}''&=& N_{1}'+\alpha t's_{p'k}, \nonumber \\
N_{2}''&=& N_{2}'-\alpha t's_{pk}, \nonumber \\
N_{4}''&=& N_{4}'-\beta t s_{q'k},  \\
N_{5}''&=& N_{5}'+\beta t s_{qk}, \nonumber \\
N_{7}''&=& N_{7}'+(\alpha+\beta) tt', \nonumber
\end{eqnarray}
is performed, where $\alpha$ and $\beta$ are rational functions of the kinematic invariants fully determined by the 
condition that the numerators $N_{i}''$ satisfy BCJ duality
\begin{eqnarray}
N_{1}''-N_{2}''+N_{7}''=0, \hspace*{1cm} N_{4}''-N_{5}''-N_{7}''=0.
\end{eqnarray}

We apply now color-kinematics duality to construct the amplitude:
\begin{eqnarray}
-i\mathcal{M}=\left({\kappa\over 2}\right)^{3}\left({N_{1}''\widetilde{N}_{1}''\over t's_{p'k}}
+{N_{2}''\widetilde{N}_{2}''\over t's_{pk}}
+{N_{4}''\widetilde{N}_{2}''\over t s_{q'k}}+{N_{5}''\widetilde{N}_{5}''\over t s_{qk}}
+{N''_{7}\widetilde{N}_{7}''\over tt'}\right),
\label{eq:cand_grav_amp}
\end{eqnarray}
that has the structure given in Eqs.~\eqref{eq:pre-tensor_structure} and \eqref{eq:tensor_structure}. The calculation
of the coefficients $\mathcal{A}_{i}$ in MRK gives
\begin{eqnarray}
\mathcal{A}_{pp}&=& \left(\alpha_{1}-2{\beta_{1}\over \beta_{2}}\right)^{2}+4\alpha_{2}\beta_{1}\left(
{\alpha_{1}\over \beta_{2}^{2}}\right)+\ldots, \nonumber \\
\mathcal{A}_{qq}&=&\left(\beta_{2}+2{\alpha_{2}\over \alpha_{1}}\right)^{2}-4\alpha_{2}\beta_{1}\left(
{\beta_{2}\over \alpha_{1}^{2}}\right)+\ldots, \nonumber \\
\mathcal{A}_{pq} &=& \left(\alpha_{1}-2{\beta_{1}\over \beta_{2}}\right)\left(\beta_{2}+2{\alpha_{2}\over \alpha_{1}}\right)
+\ldots,  
\label{}\\
\mathcal{A}_{kp} &=& -\left(\alpha_{1}-2{\beta_{1}\over \beta_{2}}\right)+\ldots, \nonumber \\
\mathcal{A}_{kq} &=& -\left(\beta_{2}+2{\alpha_{2}\over \alpha_{1}}\right)+\ldots\nonumber
\end{eqnarray}
Again, this correctly reproduces the part $\Omega^{\mu}\Omega^{\nu}$ of the MRK gravitational
amplitude, but it falls short of reproducing the terms canceling the overlapping divergences. 

To conclude our analysis, we  would like to investigate which topologies in the original Feynman diagrams contribute to the different terms in the coefficients obtained in Eq.~\eqref{eq:As_ckdual}. To trace each contribution, 
we rescale the original
numerators in Eq.~\eqref{coeficientes}  by some
constants $D_{i}$ according to $N_i \to N_i  D_i$
and study the expansion in the MRK regime. This results in the following form for the coefficients:
\begin{eqnarray}   
{\cal A}_{pp}    &=& \alpha_1^2 
       - 2 \left( \frac{D_1+D_2 }{D_7 } \right) \frac{\alpha_1 \beta_1}{\beta_2}
       + 2 \left(\frac{D_1^2+D_2^2}{D_7^2} \right)\frac{\beta_1^2}{\beta_2^2}
       +\left( \frac{D_1+D_2 }{D_7 } \right)\frac{\alpha _2 \beta _1\alpha _1 }{\beta _2^2} \nonumber\\[0.2cm]
   &-& \left[ \frac{D_1^2-3
   D_2 D_1+3 D_4 D_1+2 D_2^2+3 D_2
   D_4-4 \left(D_1+D_2\right)
   D_7 }{D_7^2 }\right] \frac{\alpha_2 \beta_1}{\beta_2}+ \dots  \nonumber\\[0.2cm]
   & &\longrightarrow \,\,\,\, \left(\alpha _1-2 \frac{D_b}{D_7}\frac{\beta _1}{\beta_2}\right)^2+\frac{\alpha _2
   \beta _1 \Big[2 D_7 
   D_b \alpha _1-\beta _2 (6 D_b^2-8 D_7
   D_b)\Big]}{D_7^2 \beta
   _2^2}.
   \end{eqnarray}
Note that, in the last line, we have simplified the expression collecting similar topologies using the same 
constant for them, {\it i.e.} $D_b \equiv D_1 = D_2 = D_4 = D_5$, where with the subscript $b$ indicates that
they are associated with the diagrams where the gluon is emitted by bremsstrahlung from a scalar line.  
$D_{3,6}$ mark the diagrams containing the 
scalar-scalar-gluon-gluon vertex and $D_7$ the diagram with the three gluon vertex.   The remaining 
results are
\begin{eqnarray}
{\cal A}_{qq}  &=& \beta_2^2 +2 \left(\frac{D_4+D_5}{D_7 }\right) \frac{\alpha_2 \beta_2}{\alpha_1} 
   + 2 \left( \frac{ D_4^2+D_5^2}{D_7^2} \right)\frac{\alpha_2^2}{\alpha_1^2}
   - \left(\frac{D_4+D_5}{D_7} \right) \frac{ \alpha _2
   \beta _1 \beta _2}{\alpha_1^2}+ \dots \nonumber\\[0.2cm]
   &+& \left[ \frac{D_4^2-3 D_5 D_4+2
   D_5^2+\left(D_4+D_5\right) \left(3
   D_1-4 D_7\right) }{D_7^2} \right] \frac{\alpha_2 \beta_1}{\alpha_1}\nonumber\\[0.2cm]
   & &\longrightarrow \,\,\,\, \left(\beta_2+2 \frac{D_b}{D_7} \frac{\alpha_2}{\alpha_1}\right)^2+\frac{\alpha _2
   \beta _1 \left(\alpha _1 \left(6
   D_b^2-8 D_7 D_b\right)-2 D_7 \beta
   _2 D_b\right)}{D_7^2 \alpha _1^2},
 \end{eqnarray}
  \begin{eqnarray}
{\cal A}_{pq} &=&    \alpha _1 \beta
   _2 -\frac{\left(D_1+D_2\right
   ) }{D_7} \beta _1 +\frac{\left(D_4+D_5\right)
   \alpha_2}{D_7} 
   -\frac{\left(D_1+D_2\right)
   \left(D_4+D_5\right) }{D_7^2} \frac{\beta _1 \alpha _2
   }{\alpha _1 \beta
   _2}+ \dots \nonumber\\[0.2cm]
   & &\longrightarrow \,\,\,\, \left(\alpha _1-2 \frac{D_b}{D_7}\frac{\beta _1}{\beta_2}\right)\left(\beta_2+2 \frac{D_b}{D_7} \frac{\alpha_2}{\alpha_1}\right),
\end{eqnarray}
\begin{eqnarray}   
{\cal A}_{kp} 
&=& -\alpha _1+  \frac{\left(D_1+D_2\right)}{D_7}\frac{\beta_1}{\beta_2} + \dots \hspace*{8cm} \nonumber\\[0.2cm]
& &\longrightarrow \,\,\,\, - \left(\alpha _1- 2  \frac{D_b}{D_7}\frac{\beta_1}{\beta_2} \right), 
   \end{eqnarray}
   \begin{eqnarray}   
{\cal A}_{kq} &=& 
-\beta _2-\frac{\left(D_4+D_5\right) }{D_7 }\frac{\alpha_2}{\alpha_1} + \dots  \hspace*{8cm} \nonumber\\[0.2cm]
& &\longrightarrow \,\,\,\, - \left(\beta _2 + 2\frac{D_b }{D_7 }\frac{\alpha_2}{\alpha_1} \right),
\end{eqnarray}
   \begin{eqnarray}   
{\cal A}_{kq} &=& 
-\beta _2-\frac{\left(D_4+D_5\right) }{D_7 }\frac{\alpha_2}{\alpha_1} + \dots   \hspace*{8cm} \nonumber\\[0.2cm]
& &\longrightarrow \,\,\,\, - \left(\beta _2 + 2\frac{D_b }{D_7 }\frac{\alpha_2}{\alpha_1} \right).
\end{eqnarray}
It is remarkable that the MRK limit is blind to the constants $D_{3,6}$ and therefore to the diagrams with the 
scalar-scalar-gluon-gluon vertex. 

\section{Discussion}

We have investigated the color-kinematics duality proposed by Bern, Carrasco, and Johannson for the 
construction of gravitational amplitudes as a formal double-copy of equivalent amplitudes in gauge theories. 
In the gauge theory side we work with scalar QCD, and study the scattering of two distinct scalar with production
of a gluon in the final state. We write the amplitude in two different representations and construct 
the corresponding ``gravitational'' scattering amplitudes using the BCJ doubling prescription.
Despite them being different, when taking the multi-Regge limit of these amplitudes 
both BCJ representations reproduce the part of the MRK gravitational vertex (with two 
reggeized gravitons and one on-shell graviton) which corresponds to the square of the MRK vertex in QCD (with 
two reggeized gluons and one on-shell gluon).  The subleading terms responsible for the fulfillment of 
the Steinmann conditions are not reproduced correctly and are dependent on the choice of numerators satisfying
BCJ duality used to write the gauge amplitude. 
This has a likely origin in the external matter states 
of the gauge amplitudes, for which the duality does not hold.

\section*{Acknowledgments} 

We thank Henrik Johansson for a useful discussion during the 2012 Informal Meeting on Scattering 
Amplitudes \& the Multi-Regge limit, at the Instituto de F\'{\i}sica Te\'orica UAM/CSIC. 
ASV acknowledges partial support from the European Comission under contract LHCPhenoNet (PITN-GA-2010-264564), 
the Madrid Regional Government through Proyecto HEPHACOS ESP-1473, the Spanish Government 
MICINN (FPA2010-17747) and Spanish MINECOÕs ``Centro de Excelencia Severo Ochoa" Programme under grant  SEV-2012-0249. The work of ESC 
has been supported by a Spanish Government FPI Predoctoral Fellowship and grant FIS2009-07238. 
MAVM acknowledges partial support from Spanish Government grants FPA2009-10612 and FIS2009-07238, 
Basque Government Grant IT-357-07 and Spanish Consolider-Ingenio 2010 Programme CPAN (CSD2007-00042).
MAVM thanks the Instituto de F\'{\i}sica Te\'orica UAM/CSIC for kind hospitality.

\end{fmffile}

\end{document}